\documentstyle[12pt]{article}

\makeatletter
\@addtoreset{equation}{section}
\makeatother

\newcommand{\bz}{{\hbox{\bf Z}}}
\newcommand{\br}{{\hbox{\bf R}}}
\newcommand{\bc}{{\hbox{\bf C}}}
\newcommand{\bq}{{\hbox{\bf Q}}}

\newcommand \bra[1]{\left< {#1} \,\right\vert}
\newcommand \ket[1]{\left\vert\, {#1} \, \right>}

\newcommand \qint[1]{\left[ {#1} \right]}

\newcommand \bosal[5]{\, a  \biggl(
{#1};{#2},{#3} \, \biggl| \, {#4} ; {#5} \biggr)\,}
\newcommand \bosbl[5]{\, b  \biggl(
{#1};{#2},{#3} \, \biggl| \, {#4} ; { #5} \biggr)\,}
\newcommand \boscl[5]{\, c  \biggl(
{#1};{#2},{#3} \, \biggl| \, {#4} ; { #5} \biggr)\,}
\newcommand \bosa[3]{\, a  \biggl(
{#1} \, \biggl| \, {#2} ; { #3} \biggr)\,}
\newcommand \bosb[3]{\, b  \biggl(
{#1} \, \biggl| \, {#2} ; { #3} \biggr)\,}
\newcommand \bosc[3]{\, c  \biggl(
{#1} \, \biggl| \, {#2} ; { #3} \biggr)\,}
\newcommand \diff[2]{{~}_{\scriptstyle {#1}}
\displaystyle \partial_{\scriptstyle {#2}} \,}
\newcommand \fra[2]{\displaystyle\frac{#1}{#2}}

\newcommand{\inp}[2]{({#1} ; {#2} )_{\infty}}
\newcommand{\jint}{\displaystyle \int_{0}^{s \infty} d_{p} t\;}
\newcommand{\hh}{\rule[-5mm]{0mm}{10mm}}
\newcommand{\hhh}{\rule[-3mm]{0mm}{7mm}}

\newcommand{\jup}{J^{-}_{I}}
\newcommand{\jdn}{J^{-}_{I\hskip -.3ex I}}
\newcommand{\jsup}{J^{S}_{I}}
\newcommand{\jsdn}{J^{S}_{I\hskip -.3ex I}}

\font\germ=eufm10 scaled \magstep1
\def\goth#1{\hbox{\germ #1}}

\def\slth{\widehat{\goth{sl}}_2\hskip 1pt}

\def\uqa{U_q\bigl(\slth\bigr)}

\def\uqap{U'_q\bigl(\slth\bigr)}

\newtheorem{prop}{Proposition}[section]

\begin{document}

\begin{flushright}
  UT-618 \\
  (Revised Version) \\
  Sept. 1992
\end{flushright}
\vspace{24pt}
\begin{center}
\begin{Large}
{\bf Free Boson Representation of $q$-Vertex Operators} \par
\vskip 3mm
{\bf and their Correlation Functions}
\end{Large}

\vspace{36pt}
Akishi Kato\raisebox{2mm}{$\star$},
Yas-Hiro Quano\raisebox{2mm}{$\star\star$} and
Jun'ichi Shiraishi

\vspace{6pt}
{\it Department of physics, University of Tokyo} \\
{\it Bunkyo-ku, Tokyo 113, Japan}

\vspace{48pt}

\underline{ABSTRACT}
\end{center}

\vspace{48pt}

A bosonization scheme of the $q$-vertex operators of $\uqa$ for
arbitrary level is obtained.  They act as intertwiners among the highest
weight modules constructed in a bosonic Fock space.  An integral formula
is proposed for $N$-point functions and explicit calculation for
two-point function is presented.

\vspace{24pt}

\vfill
\hrule

\vskip 3mm
\begin{small}

\noindent\raisebox{2mm}{$\star$} Partly supported by the Grant-in-Aid
for Scientific Research
from the Ministry of Education, Science and Culture
(No. 04245206).

\noindent\raisebox{2mm}{$\star\star$} A Fellow of the Japan
Society of the Promotion
of Science for Japanese Junior Scientists.

\noindent Partly supported by the Grant-in-Aid for Scientific Research
from the Ministry of Education, Science and Culture
(No. 04-2297).

\end{small}

\newpage

\section{Introduction}

One of the central subjects of mathematical physics has been the studies
on exactly solvable models in two dimensions for many years. Infinite
dimensional symmetries such as conformal and current algebra gives
powerful tools to investigate systems just on the critical point
\cite{ISZ}. It is now a very important problem how to extend the method
developed in the critical theories to massive field theories and lattice
models.

A breakthrough was brought by Frenkel and Reshetikhin \cite{FR} who
studied the $q$-de\-forma\-tion of the vertex operator as an intertwiner
between certain modules of quantum affine algebra $\uqa$. They showed
that the correlation functions satisfy a $q$-difference equation,
$q$-deformed Knizhnik-Zamolodchikov equation, and that the resulting
connection matrices give rise to the elliptic solution to Yang-Baxter
equation of RSOS models \cite{ABF} \cite{JMO}.  Using the $q$-vertex
operators people in Kyoto school \cite{DFJMN} succeeded in
diagonalization of the XXZ spin chain and showed that the spectra of the
XXZ model is completely determined in terms of the representation theory
of $\uqa$.  Furthermore, they found an integral formula for correlation
functions of the local operators of the XXZ model \cite{JMMN} by
utilizing bosonization of $\uqa$ of level one \cite{FJ} and the
bosonized $q$-vertex operators.

In a previous paper \cite{Sh}, one of the authors construct the
bosonization of $\uqap$ currents for arbitrary level \`{a} la Wakimoto
\cite{Wa}. In this paper we shall introduce a bosonization of the
``elementary'' $q$-vertex operators, which have exactly the same
commutation relations with the generators of $\uqap$ as the bona-fide
$q$-vertex operators have.  They are well-defined operators acting on a
bosonic Fock space, in which all the integrable highest weight modules
of a given level can be embedded.  Finally $q$-vertex operators as
intertwiners among these modules are obtained in terms of the elementary
$q$-vertex operators dressed with the screening charges.  This technique
provides a natural framework to write down an integral formula for
correlation functions of the $q$-vertex operators.  Our formula will be
useful to examine higher spin chain \cite{IIJMNT}.

The present article is organized as follows.  In section 2 we construct
the currents which give Drinfeld realization of $\uqap$ \cite{Dr} in
terms of free bosons \cite{Sh}. In section 3 we construct the
``elementary'' $q$-vertex operator.  In section 4 we define the Fock
space on which the currents and the elementary $q$-vertex operators act.
We also introduce the screening charge, which is necessary to calculate
correlation functions. Furthermore we give the expression of the
$N$-point function in terms of the bosonized operators. In section 5 we
calculate the two-point function in a simple case and show the relevance
of our formulation. In section 6 we summarize our results and give some
remark.

Three appendices are devoted to the detail of calculation in section 5.
In Appendix A OPE formulae among the bosonized operators are listed. In
Appendix B we give the normalization of the elementary vertex operators.
In Appendix C we discuss the response of Jackson integrals to $p$-shift
of valuables.

\section{Free Boson Realization of $\uqap$ }

In this section we briefly recall the bosonization of $\uqap$ \cite{Sh}.

\vskip 2mm
\noindent{\bf 2.1  Definition of $\uqa$} \quad
To begin with, let us fix notation concerning the affine Lie algebra
$\slth$ \cite{Ka}. Let $P=\bz \Lambda_{0} \oplus \bz \Lambda_{1} \oplus
\bz \delta$ be the weight lattice and $Q=\bz \alpha_{0} \oplus \bz
\alpha_{1}$ be the root lattice endowed with the symmetric bilinear form
$(~,~)$ defined by
$$
(\Lambda_{0},\Lambda_{0}) = 0 , ~
(\Lambda_{0},\alpha_{1}) = 0 , ~
(\Lambda_{0},\delta) = 1 , ~
(\alpha_{1},\alpha_{1}) = 2 , ~
(\alpha_{1},\delta) = 0 , ~
(\delta,\delta) = 0 ,
$$
where $\Lambda_{1} = \Lambda_{0}+\alpha_{1}/2 ,
\delta = \alpha_{0} + \alpha_{1} $.
As usual,
we set $\rho = \Lambda_{0} + \Lambda_{1} $.
We define
$P^{*}=\bz h_{0} \oplus \bz h_{1} \oplus \bz d$ as the dual space of $P$.
The dual pairing $\langle ~,~\rangle $ is defined by
$$
  \langle h_i, \lambda \rangle := ( \alpha_i, \lambda), \quad
  (i=0,1)\quad \mbox{for} \quad \lambda \in P.
$$

We denote by $P_{k} = \{ (k-i)\Lambda_{0} + i \Lambda_{1} |
i=0,1,\cdots,k \}$ the set of dominant integral weights of level $k$.
For simplicity, we set $\lambda_{i}= (k-i)\Lambda_{0} + i \Lambda_{1}$.

Throughout this paper let $q$ be transcendental over $\bq$ with $|q|<1$.
We use the following standard notation:
$$
\qint{m} = \fra{q^{m}-q^{-m}}{q-q^{-1}},
$$
for $m \in \bz$.

The quantum affine algebra $\uqa$ is an associative algebra over
$\bq(q)$ with 1, generated by $e_{0}, e_{1}, f_{0}, f_{1}$ and $q^{h} (h
\in P^{*})$.  The defining relations are as follows \cite{Dr2} \cite{Ji}
\begin{equation}
\begin{array}{l}
q^{h}q^{h'}=q^{h+h'}, \quad q^{0} =1, \hhh \\
q^{h}e_{i}q^{-h} = q^{ \langle h,\alpha_{i}\rangle}e_{i}, \hhh \\
q^{h}f_{i}q^{-h} = q^{-\langle h,\alpha_{i}\rangle}e_{i}, \hhh \\
\left[ e_{i}, f_{j} \right] =
\delta_{i,j} \fra{t_{i}-t_{i}^{-1}}{q-q^{-1}} \quad (t_{i}=q^{h_{i}}), \hhh \\
e_{i}^{3}e_{j} - \qint{3}e_{i}^{2}e_{j}e_{i} +
\qint{3}e_{i}e_{j}e_{i}^{2}-e_{j}e_{i}^{3} = 0 \;\; (i\neq j) ,  \hhh \\
f_{i}^{3}f_{j} - \qint{3}f_{i}^{2}f_{j}f_{i} +
\qint{3}f_{i}f_{j}f_{i}^{2}-f_{j}f_{i}^{3} = 0 \;\; (i\neq j) . \hhh
\end{array}
\end{equation}

The algebra $\uqap$ is the subalgebra of $\uqa$ generated by
$\{ e_i, f_i, t_i \;\;(i=0,1) \}$.

The algebra $\uqa$ has a Hopf algebra structure with the following
 coproduct $\Delta$ : $\uqa \rightarrow \uqa \otimes \uqa$
$$
\begin{array}{l}
\Delta (e_i)= e_i \otimes 1 + t_{i} \otimes e_i, \qquad
\Delta (f_i)= f_i \otimes t_{i}^{-1} + 1 \otimes f_i, \qquad (i=0,1)
\hhh \\
\Delta (q^{h})= q^{h} \otimes q^{h} \qquad h \in P^{*}. \hhh
\end{array}
$$

\noindent{\bf 2.2  Drinfeld Realization of $\uqap$} \quad
The Chevalley generators $e_i, f_i, t_i$ are not convenient for
considering the bosonization. We recall here the Drinfeld realization of
$\uqap$ \cite{Dr} which we will bosonize. The Drinfeld realization of
$\uqa$ is an associative algebra generated by the letters
$\{ J^{\pm}_{n} |n \in \bz \}$, $\{ J^{3}_{n} |n \in \bz_{\neq 0} \}$,
$\gamma^{\pm1/2}$ and $K$, satisfying the following relations.
\begin{equation}
\begin{array}{l}
\gamma^{\pm1/2} \in \mbox{ the center of the algebra}, \hh \\
\qint{J^{3}_{n},J^{3}_{m}} =
\delta_{n+m,0} \fra{1}{n}\qint{2n}
\fra{\gamma^{k}-\gamma^{-k}}{q-q^{-1}} , \hh \\
\qint{J^{3}_{n},K} = 0 , \hh \\
KJ^{\pm}_{n}K^{-1} = q^{\pm2}J^{\pm}_{n} , \hh \\
\qint{J^{3}_{n},J^{\pm}_{m}} =
\pm \fra{1}{n}\qint{2n} \gamma^{\mp|n|/2}J^{\pm}_{n+m} , \hh \\
J^{\pm}_{n+1}J^{\pm}_{m}-q^{\pm2}J^{\pm}_{m}J^{\pm}_{n+1} =
q^{\pm2}J^{\pm}_{n}J^{\pm}_{m+1}-J^{\pm}_{m+1}J^{\pm}_{n} , \hh \\
\qint{J^{+}_{n},J^{-}_{m}} =
\fra{1}{q-q^{-1}}(\gamma^{(n-m)/2}\psi_{n+m} -
\gamma^{(m-n)/2}\varphi_{n+m}), \hh
\end{array}
\label{Dreal}
\end{equation}
where $\{ \psi_{r}, \varphi_{s} |r, s \in \bz \}$ are related to
$\{ J^{3}_{l} | l \in \bz_{\neq 0} \}$ by
\begin{equation}
\begin{array}{l}
\displaystyle \sum_{n \in \bz} \psi_{n} z^{-n} = K
\exp \left\{ (q-q^{-1}) \sum^{\infty}_{k=1} J^{3}_{k}z^{-k} \right\} , \hh \\
\displaystyle \sum_{n \in \bz} \varphi_{n} z^{-n} = K^{-1}
\exp \left\{-(q-q^{-1}) \sum^{\infty}_{k=1} J^{3}_{-k}z^{k} \right\} . \hh
\end{array}
\end{equation}

The standard Chevalley generators $\{ e_{i},f_{i},t_{i} \}$ are given by the
identification
\begin{equation}
t_{0} = \gamma K^{-1}, \;\; t_{1} = K, \;\;
e_{1} = J^{+}_{0}, \;\; f_{1} = J^{-}_{0}, \;\;
e_{0} t_{1} = J^{-}_{1}, \;\; t^{-1}_{1} f_{0} = J^{+}_{-1} .
\end{equation}

\noindent{\bf 2.3  Bosonization of $\uqap$ } \quad
Let $k$ be a non negative integer.
Let $\{a_{n},b_{n},c_{n},Q_{a},Q_{b},Q_{c}|n \in \bz \}$
be a set of operators satisfying the following commutation relations:
\begin{equation}
\begin{array}{ll}
\qint{a_{n},a_{m}} = \delta_{n+m,0}
\fra{ \qint{2n}\qint{(k+2)n}}{ n}, &
\qint{\tilde{a}_{0},Q_{a}}=2(k+2), \hh \\
\qint{b_{n},b_{m}} = - \delta_{n+m,0}
\fra{ \qint{2n}\qint{2n}}{ n}, &
\qint{\tilde{b}_{0},Q_{b}}=-4, \hh \\
\qint{c_{n},c_{m}} = \delta_{n+m,0}
\fra{ \qint{2n}\qint{2n}}{ n}, &
\qint{\tilde{c}_{0},Q_{c}}=4, \hh
\end{array}
\end{equation}
where
$$
\begin{array}{ccc}
\tilde{a}_{0}=\fra{ q-q^{-1}}{ 2\log q} a_{0} , &
\tilde{b}_{0}=\fra{ q-q^{-1}}{ 2\log q} b_{0} , &
\tilde{c}_{0}=\fra{ q-q^{-1}}{ 2\log q} c_{0} , \hh
\end{array}
$$
and others commute.

Let us introduce the free bosonic fields $a, b,$ and $c$
carrying parameters $L,M,N \in \bz_{>0}$, $\alpha \in \br$.
Define $\bosal{L}{M}{N}{z}{\alpha}$ by
\begin{equation}
\begin{array}{rl}
\bosal{L}{M}{N}{z}{\alpha} = &
- \displaystyle \sum_{n \neq 0}
\fra{ \qint{Ln} a_{n}}
{ \qint{Mn}\qint{Nn}} z^{-n}q^{|n|\alpha}
+ \fra{ L\tilde{a}_{0}}{ MN} \log z
+ \fra{ LQ_{a}}{ MN}.
\end{array}
\end{equation}
$\bosbl{L}{M}{N}{z}{\alpha}, \boscl{L}{M}{N}{z}{\alpha}$
are defined in the same way.
In the case $L=M$ we also write
\begin{equation}
\begin{array}{rl}
\bosa{N}{z}{\alpha} =&
\bosal{L}{L}{N}{z}{\alpha} \\
=&  - \displaystyle \sum_{n \neq 0}
\fra{ a_{n}}
{ \qint{Nn}} z^{-n}q^{|n|\alpha}
+ \fra{ \tilde{a}_{0}}{ N} \log z
+ \fra{ Q_{a}}{ N},
\end{array}
\end{equation}
and likewise for $\bosb{N}{z}{\alpha}, \bosc{N}{z}{\alpha}$.

Let
$\{a_{n},b_{n},c_{n}|n \in \bz_{\geq 0} \}$ be annihilation operators,
and $\{a_{n},b_{n},c_{n},Q_{a},Q_{b},Q_{c}|n \in \bz_{< 0} \}$ creation
operators.
We denote by $: {\cal O}(z) :$ the normal ordering of ${\cal O}(z)$.
For example,
$$
:\exp \left\{ \bosb{2}{z}{\alpha} \right\}: =
\exp \left\{- \displaystyle \sum_{n < 0}
\fra{b_{n}}{\qint{2n}} z^{-n} q^{|n|\ \alpha} \right\}
\exp \left\{- \displaystyle \sum_{n > 0}
\fra{b_{n}}{\qint{2n}} z^{-n} q^{|n|\ \alpha} \right\}
e^{Q_{b}/2} z^{\tilde{b}_{0}/2}.
$$

Now we define
the currents $J^{3}(z), J^{\pm}(z)$
as follows:
\begin{equation}
\begin{array}{rl}
J^{3}(z) =& \diff{k+2}{z} \bosa{k+2}{q^{-2}z}{-1}
+\diff{2}{z} \bosb{2}{q^{-k-2}z}{-{ \fra{k+2}{2}}} , \hh \\
J^{+}(z) =&
-:\left[\diff{1}{z} \exp \left\{ -\bosc{2}{q^{-k-2}z}{0} \right\} \right]
\times \exp
\left\{ -\bosb{2}{q^{-k-2}z}{1} \right\}: , \hh \\
J^{-}(z) =&
:\left[ \diff{k+2}{z} \exp \left\{
\bosa{k+2}{q^{-2}z}{-{\fra{k+2}{2}}}
+\bosb{2}{q^{-k-2}z}{- 1} \right. \right. \hh \\
& \hspace{2.5cm} \left. \left.
+\boscl{k+1}{2}{k+2}{q^{-k-2}z}{0} \right\}\right] \hh \\
& \times \exp \left\{-\bosa{k+2}{q^{-2}z}{\fra{k+2}{2}}
+\boscl{1}{2}{k+2}{q^{-k-2}z}{0} \right\}: . \hh
\end{array}
\label{Jdef}
\end{equation}
Here the $q$-difference operator with parameter $n \in \bz_{>0}$ is
defined by
$$
\diff{n}{z} f(z) \equiv \fra{ f(q^{n}z) - f(q^{-n}z)}
{ (q-q^{-1})z}.
$$

Define further the
auxiliary fields $\psi(z), \varphi(z)$ as
\begin{equation}
\begin{array}{rl}
\psi(z) =& :\exp \left\{
(q-q^{-1}) \displaystyle \sum_{n>0} (q^{n} a_{n}+ q^{(k+2)n/2} b_{n}) z^{-n}
+ (\tilde{a}_{0}+\tilde{b}_{0}) \log q \right\}: , \hh \\
\varphi(z) =& :\exp \left\{ -(q-q^{-1}) \displaystyle \sum_{n<0}
(q^{3n}a_{n}+q^{3(k+2)n/2}b_{n}) z^{-n}
- (\tilde{a}_{0}+\tilde{b}_{0}) \log q \right\}: . \hh
\end{array}
\label{psidef}
\end{equation}

We give the mode expansions of these fields as
\begin{equation}
\begin{array}{rlrl}
\displaystyle \sum_{n \in \bz} J^{3}_{n} z^{-n-1} =& J^{3}(z) , &
\displaystyle \sum_{n \in \bz} J^{\pm}_{n} z^{-n-1} =& J^{\pm}(z) , \\
\displaystyle \sum_{n \in \bz} \psi_{n} z^{-n} =& \psi(z) , &
\displaystyle \sum_{n \in \bz} \varphi_{n} z^{-n} =& \varphi(z).
\end{array}
\label{mode}
\end{equation}
and let
\begin{equation}
   K= q^{\tilde{a}_{0}+\tilde{b}_{0}}, \qquad \gamma = q^{k}.
  \label{Kdef}
\end{equation}

Then we get the following \cite{Sh}:
\begin{prop}
$ \{ J^{3}_{n} | n \in \bz_{\neq 0} \}$, $ \{ J^{\pm}_{n} | n \in \bz
\}$ $ \{ \varphi_{n}, \psi_{n} | n \in \bz \}$, $K$, and $\gamma$
defined by (\ref{Jdef}), (\ref{psidef}), (\ref{mode}) and (\ref{Kdef})
satisfy the relations (\ref{Dreal}).
\end{prop}

\noindent{\bf 2.4  Finite Dimensional $\uqa$ Module} \quad
For $l\in \bz_{\ge 0}$ let $V^{(l)}$ denote the $(l+1)$-dimensional
$U'_q\bigl(\slth\bigr)$-module (spin $l/2$ representation) with basis
$\{v^{(l)}_{m} | 0\le m\le l\}$ given by
\begin{eqnarray*}
   &&e_{1}v^{(l)}_{m}=[m]v^{(l)}_{m-1},
   \quad
   f_1v^{(l)}_m=[l-m]v^{(l)}_{m+1},
   \quad
    t_1v^{(l)}_m=q^{l-2m}v^{(l)}_{m},\\
   &&
   e_0=f_1,
   \quad f_0=e_1,
   \quad t_0=t_1^{-1}
   \quad \mbox{on}\quad V^{(l)}.
\end{eqnarray*}

Here $v^{(l)}_{m}$ with $m<0$ or $m>l$ is understood to be $0$.
In the case $l=1$ we also write $v^{(1)}_0=v_+$ and $v^{(1)}_1=v_-$.

We equip $V_z^{(l)}=V^{(l)}\otimes \bq(q)[z,z^{-1}]$ with a
$\uqa$-module structure via
\begin{eqnarray*}
   &&
   e_i(v^{(l)}_{m}\otimes z^n)=e_iv^{(l)}_{m}\otimes z^{n+{\delta_{i\,0}}},
   \quad
   f_i(v^{(l)}_{m}\otimes z^n)=f_iv^{(l)}_{m}\otimes z^{n-\delta_{i\,0}},
   \\
   &&
   {\rm wt}(v^{(l)}_{m}\otimes z^n)=n\delta+ (l-2m)(\Lambda_{1}- \Lambda_{0})
\end{eqnarray*}
Namely $V_z^{(l)}$ is the affinization of $V^{(l)}$

We also need the representation of Drinfeld generators on level 0 modules.

\begin{prop}\label{Drin}
Spin $l/2$ representation of $\uqa$
is given in terms of the Drinfeld generators by
\begin{eqnarray}
    \gamma^{\pm 1/2} v^{(l)}_{m} &=& v^{(l)}_{m}, \nonumber \\
    K v^{(l)}_{m} &=& q^{l-2m} v^{(l)}_{m}, \nonumber \\
    J^{+}_{n} v^{(l)}_{m} &=& z^{n} q^{n(l-2m+2)} [l-m+1] v^{(l)}_{m-1}, \\
    J^{-}_{n} v^{(l)}_{m} &=& z^{n} q^{n(l-2m)} [m+1] v^{(l)}_{m+1},
    \nonumber \\
    J^{3}_{n} v^{(l)}_{m} &=& \fra{z^{n}}{n} \left\{
           [nl] - q^{n(l+1-m)} (q^{n}+q^{-n}) [nm] \right\},
    \nonumber
\end{eqnarray}
where $v^{(l)}_{m}=0$ if $m>l$ or $m<0$.
\end{prop}

\section{Elementary $q$-Vertex Operators}
In this section we construct the operators which have exactly the same
commutation relations with the generators of $\uqap$ as the bona-fide
$q$-vertex operators have.

A vector $\ket{\lambda}$ is called a highest weight vector of weight
$\lambda$ if it satisfies the highest weight condition\footnote{ We do
not impose the irreducibility conditions $ f_{i}^{\langle h_{i},\lambda
\rangle + 1} \ket{\lambda} = 0, \quad i=0,1$.}
$$
   e_{i} \ket{\lambda} = 0,
   \quad t_{i} \ket{\lambda} = q^{\langle h_{i},\lambda \rangle}
  \ket{\lambda}, \quad i=0,1.
$$
The left highest weight module $V(\lambda)$ with the highest weight
vector $\ket{\lambda}$ is defined by
$$
V(\lambda):=\uqa \ket{\lambda}.
$$
The right highest weight module is defined
in a similar manner.

The left (resp. right) highest weight module with highest weight
$\lambda\in P_k$ will be denoted by $V(\lambda)$ (resp. $V^r(\lambda)$).
We fix a highest weight vector $\ket{\lambda}\in V(\lambda)$ (resp.
$\bra{\lambda} \in V^r(\lambda)$) once for all.  There is a unique
symmetric bilinear pairing $V^r(\lambda)\times V(\lambda)\rightarrow F$
such that
\begin{eqnarray*}
   \lefteqn{\bra{\lambda}\lambda\rangle=1,
   \qquad
   \bra{ux}u'\rangle=\bra{u}xu'\rangle }
   \\
   &&
   \forall x \in \uqa,
   \quad \forall \bra{u}\in V^r(\lambda),
   \quad \forall \ket{u'}\in V(\lambda).
\end{eqnarray*}

\noindent{\bf 3.1  Definition of $q$-Vertex Operators} \quad
We recall below the properties of the $q$-vertex operators ($q$-VOs)
relevant to the subsequent discussions. For more detail, see
\cite{DFJMN}\cite{DJO}.

Fix positive integers $k$, $l$ and let $\lambda, \mu \in P_{k}$.
We set $\Delta_{\lambda}=(\lambda,\lambda+2\rho)/2(k+2).$

We shall use the following type of VO\footnote{This VO is called ``type
I'' in ref. \cite{DFJMN}}
\begin{eqnarray}
   \Phi_{\lambda}^{\mu V^{(l)}} (z) &=& z^{\Delta_{\mu} - \Delta_{\lambda}}
   \tilde{\Phi}_{\lambda}^{\mu V^{(l)}} (z) \nonumber \\
   \tilde{\Phi}_{\lambda}^{\mu V^{(l)}} (z) &:&
   V(\lambda) \longrightarrow V(\mu)\widehat{\otimes}V_{z}^{(l)}
   \label{VO1}
\end{eqnarray}
The map (\ref{VO1}) means a formal series of the form
\begin{eqnarray*}
 &&  \tilde{\Phi}_{\lambda}^{\mu V^{(l)}} (z) = \sum_{n \in \bz}
   \sum_{m=0}^{l}
   \tilde{\Phi}_{m,n}
   \otimes
   v_{m}^{(l)} z^{-n} \\
 &&  \tilde{\Phi}_{m,n} : V(\lambda)_{\nu} \longrightarrow
   V(\mu)_{\nu-{\rm wt}(v^{(l)}_{m}) + n\delta}
\end{eqnarray*}
where ${\rm wt}(v^{(l)}_{m}) = (l-2m)(\Lambda_{1}-\Lambda_{0})$,
$\delta=\alpha_{0}+\alpha_{1}$.

By definition, the $q$-VO satisfies the intertwining relations
\begin{equation}
   \tilde{\Phi}_{\lambda}^{\mu V^{(l)}} (z) \circ x =
   \Delta (x) \circ \tilde{\Phi}_{\lambda}^{\mu V^{(l)}} (z),
   \qquad \forall x \in \uqa.
   \label{intertwine}
\end{equation}

{}From the general arguments on $q$-VOs \cite{DJO}, in our case there
exists at most one VO up to proportionality. We normalize $
\tilde{\Phi}_{\lambda}^{\mu V^{(l)}} (z) $ such that the leading term is
$\ket{\mu} \otimes v_{m}^{(l)} $ :
\begin{equation}
   \tilde{\Phi}_{\lambda}^{\mu V^{(l)}} (z) \ket{\lambda} =
   \ket{\mu} \otimes v_{m}^{(l)} + \cdots,
   \label{norm}
\end{equation}
where $\cdots$ means terms of the form $u \otimes v, {\rm wt}~u \neq \mu$.

\begin{prop}\label{VOrecurr}
If we write
$$
   \tilde{\Phi}_{\lambda}^{\mu V^{(l)}} (z)
   = \sum_{m=0}^{l}    \tilde{\Phi}_{\lambda m}^{\mu V^{(l)}} (z)
  \otimes v^{(l)}_{m},
$$
then
\begin{equation}
   \tilde{\Phi}_{\lambda m-1}^{\mu V^{(l)}} (z)
   = \frac{1}{[l-m]}
   \left\{
   \tilde{\Phi}_{\lambda m}^{\mu V^{(l)}} (z)
   f_{1} - q^{2m-l} f_{1}
   \tilde{\Phi}_{\lambda m}^{\mu V^{(l)}} (z)
   \right\}, \qquad m=1,2,\ldots,l.
\end{equation}
\end{prop}
This is easily checked by evaluating the both sides of
(\ref{intertwine}) for $x=f_{1}$.

For two vertex operators
\begin{eqnarray*}
 && \tilde{\Phi}_{\mu_{0}}^{\mu_{1} V^{(l_{1})}} (z_{1}) = \sum_{n \in
\bz} \sum_{m=0}^{l} \tilde{\Phi}_{m,n} \otimes v_{m}^{(l_{1})} z_1^{-n}
, \\ && \tilde{\Phi}_{\mu_{1}}^{\mu_{2} V^{(l_{2})}} (z_{2}) = \sum_{n
\in \bz} \sum_{m=0}^{l} \tilde{\Phi}_{m,n} \otimes v_{m}^{(l_{2})}
z_2^{-n} ,
\end{eqnarray*}
the composition of these two is defined as a formal series in
$z_{1}, z_{2}$ :
$$
  \tilde{\Phi}_{\mu_{1}}^{\mu_{2} V^{(l_{2})}} (z_{2})
  \circ
  \tilde{\Phi}_{\mu_{0}}^{\mu_{1} V^{(l_{1})}} (z_{1})
  = \sum \tilde{\Phi}_{jm} \circ \tilde{\Phi}_{kn}
    \otimes v^{(l_2)}_{j}z_{2}^{-m}
    \otimes v^{(l_1)}_{k}z_{1}^{-n}.
$$
The composition of $N$ $q$-vertex operators are defined in a similar
fashion.

\vskip 2mm
\noindent{\bf 3.2  Elementary $q$-Vertex Operators} \quad
In \cite{JMMN} an integral formula for correlation functions of the
local operators of the XXZ model is obtained by utilizing bosonization
of the $\uqa$ of level one \cite{FJ} and the bosonized $q$-vertex
operators.  In the same spirit we derive the formulae for the $q$-vertex
operators for arbitrary level $k$ in terms of bosonic fields $a$, $b$
and $c$.

Since the Drinfeld generators are successfully bosonized, we want to
know how the intertwining properties are expressed in those terms
\cite{CP}.
\begin{prop}\label{CPcopro}
For $k \in \bz_{\geq0}$ and $l \in \bz_{>0}$ we have
$$
   \begin{array}{l} \left.
   \begin{array}{lcl} \Delta(J_{k}^{+}) &=&
   J_{k}^{+}\otimes\gamma^{k}+\gamma^{2k}K\otimes J_k^{+}+ \displaystyle
   \sum_{i=0}^{k-1}\gamma^{(k+3i)/2}\psi_{k-i}\otimes\gamma^{k-i}J_i^{+}
   \\
   \Delta(J_{-l}^{+})&=& J_{-l}^{+}\otimes\gamma^{-l}+K^{-1}\otimes
   J_{-l}^{+}+ \displaystyle \sum_{i=1}^{l-1}\gamma^{(l-i)/2}\varphi_{-l+i}
   \otimes\gamma^{-l+i}J_{-i}^{+}
   \end{array} \right\} \bmod N_{-}\otimes
   N_{+}^2,
   \hh
   \end{array}
$$

$$
   \begin{array}{l}
   \left.  \begin{array}{lcl} \Delta(J_{l}^{-}) &=&
   J_{l}^{-}\otimes K + \gamma^{l}\otimes J_{l}^{-} + \displaystyle
   \sum_{i=1}^{l-1}\gamma^{l-i}J_{i}^{-}\otimes\gamma^{(i-l)/2}\psi_{l-i}
   \\ \Delta(J_{-k}^{-}) &=& J_{-k}^{-}\otimes\gamma^{-2k} K^{-1} +
   \gamma^{-k}K\otimes J_{-k}^{-}+ \displaystyle
   \sum_{i=0}^{k-1}\gamma^{i-k}J_{-i}^{-}
   \otimes\gamma^{-(k+3i)/2}\varphi_{i-k} \end{array} \right\} \bmod
   N_{-}^2\otimes N_{+}, \hh \\ \left.  \begin{array}{lcl}
   \Delta(J^{3}_l)&=&J^{3}_l\otimes \gamma^{l/2}+\gamma^{3l/2}\otimes
   J^{3}_l \hh \\ \Delta(J^{3}_{-l})&=&J^{3}_{-l}\otimes\gamma^{-3l/2}
   +\gamma^{-l/2}\otimes J^{3}_{-l} \hh \end{array} \right\} \bmod
   N_{-}\otimes N_{+}.
   \end{array}
$$
Here $N_{\pm}$ and $N_{\pm}^2$ are left
${\bf Q}(q)[\gamma^\pm,\psi_r,\varphi_s|r,-s\in\bz_{\geq0}]$-modules
 generated by $\{J_m^{\pm}| m \in \bz \}$ and
$\{J_m^{\pm}J_n^{\pm}|m,n\in \bz\}$ respectively.
\end{prop}

By using Propositions \ref{Drin}, \ref{CPcopro} and noting that $N_{+}
v^{(l)}_{0}=N_{-} v^{(l)}_{l}=0$, $N_{\pm}v^{(l)}_{m}\subset
F[z,z^{-1}]v^{(l)}_{m\mp 1}$, we get the exact relations
\begin{eqnarray}
   \hbox{$[J^{3}_n,\tilde{\Phi}_{\lambda l}^{\mu V^{(l)}}(z)]$}&=&
   q^{2n} z^{n} \frac{[nl]}{n}
   \cdot q^{k(n+|n|/2)}
   \tilde{\Phi}_{\lambda l}^{\mu V^{(l)}}(z) \qquad n \neq 0,
   \nonumber
\\
   \hbox{$[\tilde{\Phi}_{\lambda l}^{\mu V^{(l)}}(z),J^{+}(w)]$}&=&0,
   \label{comm}
\\
   K \tilde{\Phi}_{\lambda l}^{\mu V^{(l)}}(z) K^{-1}
    &=& q^{l}\tilde{\Phi}_{\lambda l}^{\mu V^{(l)}}(z),
   \nonumber
\end{eqnarray}
which follows from $V(\mu) \otimes v^{(l)}_{l}$ components
of intertwining relation.

These conditions put stringent constraints on the possible bosonized
form $\phi^{(l)}_{l}(z)$ of vertex operators $\tilde{\Phi}_{\lambda
l}^{\mu V^{(l)}}(z)$.  By the explicit calculation, we can check that if
\begin{equation}
   \phi^{(l)}_{l}(z)= ~~: \exp \left\{
    \bosal{l}{2}{k+2}{q^{k}z} {\fra{k+2}{2}} \right\} :
   \label{eVOdef1}
\end{equation}
is substituted for $\tilde{\Phi}_{\lambda l}^{\mu V^{(l)}}(z)$, then all
the commutation relations (\ref{comm}) hold.  Proposition
\ref{VOrecurr} suggest that the other components of
the vertex operator should be defined by the following multiple contour
integral
\begin{eqnarray}
   \lefteqn{
   \phi_{m}^{(l)} (z) =
   \frac{1}{[1][2]\cdots[l-m]}
   \oint dw_{1}
   \oint dw_{2}
   \cdots
   \oint dw_{l-m}
   }
   \nonumber
   \\
   && \qquad \qquad \times
   [\cdots[~[ \phi^{(l)}_{l}(z), J^{-}(w_{1}) ]_{q}, J^{-}(w_{2}) ]_{q^{2}}
    \cdots  J^{-}(w_{l-m}) ]_{q^{l-m}} .
   \label{eVOdef2}
\end{eqnarray}

We will call these operators (\ref{eVOdef1}), (\ref{eVOdef2}) as
``elementary vertex operators''.  A salient feature of these operators
is that they are determined solely from the commutation relation with
bosonized $\uqa$ currents; this is completely independent of which
infinite dimensional modules they intertwine\footnote{ These elementary
$q$-vertex operators are determined from a part of the intertwining
properties, but it is very likely that they enjoy all of these
properties.  }.

Before discussing the relation between ``elementary $q$-vertex
operators'' and bona-fide vertex operators, we need to clarify on which
space these bosonized operators are acting.

\section{Fock Module, Screening Charge and Correlation \break Function}

In this section, we define the Fock module of bosons on which the $\uqa$
currents $J^{3}(z), J^{\pm}(z)$, and the elementary $q$-vertex operators
$\phi^{(l)}_{m}(z)$ act. All the integrable highest weight modules are
constructed in this Fock module. Further the $q$-vertex operators as the
intertwiner among these modules are obtained.

\vskip 2mm
\noindent{\bf 4.1 Fock module and Highest Weight Module} \quad
{}From the observation that
\begin{equation}
\qint{ J^{3}(z) , \tilde{b}_{0}+\tilde{c}_{0} } = 0 , \qquad
\qint{ J^{\pm}(z) , \tilde{b}_{0}+\tilde{c}_{0} } = 0 , \qquad
\qint{ \phi^{(l)}_{m}(z) , \tilde{b}_{0}+\tilde{c}_{0} } = 0 ,
\label{b+c}
\end{equation}
we can restrict the full Fock module of the boson $a, b,$ and $c$ to the
sector such that the eigen\-value of the operator
$\tilde{b}_{0}+\tilde{c}_{0}$ is equal to 0. This requirement does not
conflict with any other conditions we shall impose.\footnote{ This kind
of decoupling is well known in CFT when we bosonize fermionic ghosts
\cite{FMS}.}

Let us introduce a vacuum vector $\ket{0}$ which has the following properties:
$$
\begin{array}{llll}
a_{n} \ket{0} =0, & b_{n} \ket{0} =0, & c_{n} \ket{0} =0, & n \geq 0. \hh
\end{array}
$$
Define the vectors $\ket{r,s}$ by
\begin{equation}
\ket{r,s} :=
\exp \left\{ r\fra{Q_{a}}{k+2}+s\fra{Q_{b}+Q_{c}}{2} \right\} \ket{0} , \hh
\end{equation}
where $ r \in \frac{1}{2}\bz , s \in \bz $.

Let $F$ be a free $\bq(q)$ module generated by $\{ a_{-1},a_{-2},\cdots,
b_{-1},b_{-2},\cdots,c_{-1},c_{-2},\cdots \}$. Now we define the Fock
modules $F_{r,s}$ as
$$
F_{r,s} := F \ket{r,s} . \hh
$$

We can regard the currents $J^{3}(z), J^{\pm}(z), J^{S}(z)$, and
$q$-vertex operators $\phi^{(l)}_{m}(z)$ as the following maps:
\begin{equation}
\begin{array}{lllll}
J^{3}(z) &: &F_{r,s} &\rightarrow &F_{r,s} , \hhh \\
J^{\pm}(z) &: &F_{r,s} &\rightarrow &F_{r,s \mp 1} , \hhh \\
\phi^{(l)}_{m}(z) &: &F_{r,s} &\rightarrow &F_{r+l/2,s+l-m} , \hhh \\
\end{array}
\end{equation}

We can check that $\ket{i/2,0}$ satisfies the highest weight
condition
$$
\begin{array}{l}
t_1 \ket{i/2,0} = q^{i} \ket{i/2,0} , \quad
t_0 \ket{i/2,0} = q^{k-i} \ket{i/2,0} , \quad
e_{0} \ket{i/2,0} = 0 , \quad
e_{1} \ket{i/2,0} = 0 . \hh
\end{array}
$$

Thus we can identify
$$
\ket{\lambda_{i}} = \ket{i/2,0}.
$$

We construct the left highest weight representations $V(\lambda_i)$
of $\uqa$ as follows\footnote{
As is well known in CFT, $f_{1}^{2i+1} \ket{i/2,0} = 0$ but
$ f_{0}^{k-2i+1} \ket{i/2,0} \neq 0$. So our module is reducible.}
$$
   V(\lambda_i) := \uqa \ket{\lambda_i}.
$$

\begin{prop}
Using this highest weight vector, we can embed the left highest weight module
$V(\lambda_{i})$ in the Fock modules as follows:
\begin{equation}
V(\lambda_{i}) \hookrightarrow
\bigoplus_{s \in \bz} F_{i/2,s} .
\label{inFock}
\end{equation}
\end{prop}

We can not simply use the vector
$\ket{r,s} , s \neq 0, $
as the highest weight vector since $e_{1} \ket{r,s}$ does not
vanish.

\vskip 2mm
\noindent{\bf 4.2 Screening Charge} \quad
We see that due to nontrivial charge assignment, naive composition of
elementary vertex operators does not define a map between highest weight
modules defined in the previous section.  This conundrum is solved by
introducing screening charge.

Let us define the screening operator $J^{S}(z)$ as follows \cite{Sh}:
\begin{equation}
\begin{array}{rl}
J^{S}(z) =& - :\left[ \diff{1}{z}
\exp \left\{ -\bosc{2}{q^{-k-2}z}{0} \right\} \right] \hh \\
& \times \exp \left\{ -\bosb{2}{q^{-k-2}z}{-1}
- \bosa{k+2}{q^{-2}z}{-\fra{k+2}{2}} \right\} : . \hh
\end{array}
\end{equation}
Then we get the following:
\begin{equation}
\begin{array}{l}
\qint{ J^{3}_{n} , J^{S}(z) } = 0 , \hhh \\
\qint{ J^{+}_{n} , J^{S}(z) } = 0 , \hhh \\
\qint{ J^{-}_{n} , J^{S}(z) } = \diff{k+2}{z} \left[ z^{n}
:\exp \left\{ -\bosa{k+2}{q^{-2}z}{\fra{k+2}{2}}
\right\}: \right], \hhh
\end{array}
\label{JSope}
\end{equation}
for all $n \in \bz$.

For $p \in \bc$, $|p|<1$, and $s \in \bc^{\times}$,
the Jackson integral is defined as
$$
\jint f(t) = s(1-p) \displaystyle
\sum^{\infty}_{m=-\infty} f(sp^{m})p^{m} ,
$$
whenever the RHS converges \cite{Ma1}.

Note that the RHS of (\ref{JSope}) is a total $p=q^{2(k+2)}$ difference.
Therefore, the following Jackson integral of the screening operator
(screening charge)
\begin{equation}
\jint J^{S}(t)
\end{equation}
commutes with all the generators of
$\uqa$ exactly.

The screening operator enjoys the same relations
\begin{equation}
\qint{ J^{S}(z) , \tilde{b}_{0}+\tilde{c}_{0} } = 0 ,
\end{equation}
as (\ref{b+c}), and is a map among Fock modules as follows
\begin{equation}
 J^{S}(z) : F_{r,s} \rightarrow F_{r-1,s-1}.
\end{equation}

We want to construct a $\uqa$-homomorphism $ V(\lambda) \rightarrow
V(\mu) \otimes V^{(l)}_{z}$. Let us consider the following combination
of operators
$$
   J^{S}(t_{1})
   J^{S}(t_{2})
   \cdots
   J^{S}(t_{l-m})
   \phi^{(l)}_{m}(z)
   \; : \; F_{r,s} \rightarrow F_{r-l/2+m,s} .
$$
By performing Jackson integral of this operator we obtain a
$\uqa$-linear map
$$
   \sum_{m}
   \int_{0}^{s_{1}\infty} d_{p} t_{1}
   J^{S}(t_{1})
   \cdots
   \int_{0}^{s_{l-m}\infty} d_{p} t_{l-m} \\
   J^{S}(t_{l-m})
   \phi^{(l)}_{m}(z) \otimes v^{(l)}_{m}
   \; : \; V(\lambda) \rightarrow V(\mu) \otimes V^{(l)}_{z},
$$
for arbitrary $\lambda, \mu \in P_{k}$.  Note that this operator depends
on $k$ and $l$ but is independent of the choice of $\lambda, \mu$.
Since we fixed the normalization of $q$-vertex operator in (\ref{norm}),
we have to choose an appropriate normalization factor for each $\lambda,
\mu \in P_k$ and $V^{(l)}$.

Now we are in a position to state our main proposition :
\begin{prop}
The $q$-vertex operator is bosonized as
\begin{equation}
   \tilde{\Phi}_{\lambda}^{\mu V^{(l)}} (z)
   = \sum_{m=0}^{l}
   \tilde{\Phi}_{\lambda m}^{\mu V^{(l)}} (z)
   \otimes v^{(l)}_{m},
\end{equation}
where
\begin{equation}
   \tilde{\Phi}^{\mu V^{(l)}}_{\lambda m}(z)
   =  g^{\mu V^{(l)}}_{\lambda}(z)
   \int_{0}^{s_{1}\infty} d_{p} t_{1}
   J^{S}(t_{1})
   \cdots
   \int_{0}^{s_{l-m}\infty} d_{p} t_{l-m} \\
   J^{S}(t_{l-m})
   \phi^{(l)}_{m}(z),
\end{equation}
and $g^{\mu V^{(l)}}_{\lambda}(z)$
is the normalization factor mentioned above.
\end{prop}

$N$-point function of the $q$-vertex operators is by definition the
expectation value of the composition
\begin{equation}
   \Phi^{\mu_{N} V_{N}}_{\mu_{N-1}} (z_{N})
   \circ
   \cdots
   \circ
   \Phi^{\mu_{1} V_{1}}_{\mu_{0}} (z_{1})
   :
   V(\mu_{0}) \rightarrow V(\mu_{N})
   \otimes V^{(l_{N})}_{z_{N}}
   \otimes
   \cdots
   \otimes V^{(l_{1})}_{z_{1}}.
\end{equation}
As a corollary of Proposition 4.2 we have
\begin{prop}
If we expand $N$-point function of $q$-vertex operators as
\begin{eqnarray*}
   \lefteqn{
   \bra{\mu_{N}}
   \Phi^{\mu_{N} V_{N}}_{\mu_{N-1}} (z_{N})
   \circ
   \cdots
   \circ
   \Phi^{\mu_{1} V_{1}}_{\mu_{0}} (z_{1})
   \ket{\mu_{0}}
   } \\
&& =
   \sum_{m_{1},\ldots,m_{N}}
   f_{m_{1},\ldots,m_{N}}(z_{1},\cdots,z_{N})
   ~~
   v^{(l_{N})}_{m_{N}}
   \otimes
   \cdots
   \otimes
   v^{(l_{1})}_{m_{1}}
   \quad \in V^{(l_{N})} \otimes \cdots \otimes  V^{(l_{N})},
\end{eqnarray*}
where $\mu_{0},\cdots,\mu_{N} \in P_{k}$,
then each component has the following integral form
\begin{eqnarray}
\lefteqn{
      f_{m_{1},\ldots,m_{N}}(z_{1},\cdots,z_{N}) =
    \prod_{i=1}^{N} z_{i}^{\Delta_{\mu_i}-\Delta_{\mu_{i-1}}}
    g_{\mu_{i-1}}^{\mu_{i} V_{i}} (z_{i}) \times} \nonumber \\
&& \times
   \bra{\mu_{N}}
     \int_{0}^{s^{(N)}_1\infty} d_{p} t^{(N)}_{1} J^S(t^{(N)}_{1})
   \cdots
   \int_{0}^{s^{(N)}_{l_{N}-m_{N}}\infty} d_{p} t^{(N)}_{l_{N}-m_{N}}
   J^S(t^{(N)}_{l_{N}-m_{N}})
   \phi^{(l_{N})}_{m_{N}} (z_{N}) \nonumber \\
&& \qquad \times
   \cdots\cdots \nonumber \\
&& \qquad \times
   \int_{0}^{s^{(1)}_1\infty} d_{p} t^{(1)}_{1} J^S(t^{(1)}_{1})
   \cdots
   \int_{0}^{s^{(1)}_{l_{1}-m_{1}}\infty} d_{p} t^{(1)}_{l_{1}-m_{1}}
   J^S(t^{(1)}_{l_{1}-m_{1}})
   \phi^{(l_{1})}_{m_{1}} (z_{1})
   \ket{\mu_{0}}.
\end{eqnarray}
\end{prop}

\section{Calculation of Two-Point Function}

In what follows we denote
$V:=V^{(1)}=\bc v_{+} \oplus \bc v_{-}$,
and $z:=z_1 /z_2 $, for short.
Let $\Psi (z_1 , z_2 )\in V \otimes V \otimes z^{3/4(k+2)}\bq (q)[[z]]$
be the following two-point function
\begin{equation}
\Psi (z_1 , z_2 ):=\bra{\lambda _0 }
\Phi_{\lambda _1 }^{\lambda _0 V}(z_2 ) \circ
\Phi_{\lambda _0 }^{\lambda _1 V}(z_1 ) \ket{\lambda _0 },
\end{equation}
where,
$$
\Phi_{\lambda _0 }^{\lambda _1 V}(z_1 )=
z^{3/4(k+2)}\tilde{\Phi}_{\lambda _0 }^{\lambda _1 V}(z_1 ), \qquad
\Phi_{\lambda _1 }^{\lambda _0 V}(z_2 )=
z^{-3/4(k+2)}
\tilde{\Phi}_{\lambda _1 }^{\lambda _0 V}(z_2 ).
$$

In this section by evaluating this correlation function,
we prove Proposition 4.3 for $N=2, l_1 =l_2 =1$, and $k\in \bz _{> 0}$.

\vskip 2mm
\noindent{\bf 5.1 Jackson Integral Formula for Two-Point Function} \quad
{}From Proposition 4.2 $q$-VOs have the following bosonization
\begin{eqnarray}
\tilde{\Phi}_{\lambda _0 }^{\lambda _1 V}(z_1 ) &=&
g_{\lambda_0 }^{\lambda _1 V}(z_1 ) \left(
\phi_+ (z_1 )\otimes v_+ +
\phi_- (z_1 )\otimes v_-
\right) ,
\\
\tilde{\Phi}_{\lambda _1 }^{\lambda _0 V}(z_2 ) &=&
g_{\lambda_1 }^{\lambda _0 V}(z_2 )
\jint J^{S}(t)
\left(
\phi_+ (z_2 )\otimes v_+ +
\phi_- (z_2 )\otimes v_-
\right) ,
\end{eqnarray}
where, $\phi_+ (z_i ) = \phi^{(1)}_{0}(z_{i})$, and
$\phi_- (z_i ) = \phi^{(1)}_{1}(z_{i}) \quad (i=1,2)$.

Here
$g_{\lambda_0 }^{\lambda _1 V}(z_1 ) =1$ and
$g_{\lambda_1 }^{\lambda _0 V}(z_2 ) =: g(z_2)$ are
 the normalization factors of
$\tilde{\Phi}_{\lambda _0 }^{\lambda _1 V}(z_1 )$ and
$\tilde{\Phi}_{\lambda _1 }^{\lambda _0 V}(z_2 )$, respectively.
Explicitly,
(as for detail calculation, see Appendix B.)
\begin{equation}
g(z_2 )=-q^{-2-(k+8)/2(k+2)}z_{2}^{-1/2(k+2)}
\left[ \jint t^{-1-2/(k+2)}
\fra{\inp{p^2 qz_2 /t}{p}}{\inp{pq^{-1}z_2 /t}{p}}\right]^{-1}
\!\!\!\!\!\!\!,
\end{equation}
where, $p=q^{2(k+2)}$, and
$$
\inp{a}{q}:=\prod_{n=0}^{\infty} (1-aq^n ).
$$

Since the vertex operators preserve the weight modulo $\delta $ we have
$$
\Psi (z_1 , z_2 )=f_1 (z_1 , z_2)v_{+}\otimes v_{-}
+f_2 (z_1 , z_2)v_{-}\otimes v_{+}.
$$

Using the free boson representation of the $q$-vertex operators
we can rewrite $f_1 (z_1 , z_2)$ as
\begin{eqnarray}
f_1 (z_1 , z_2)
& = & z^{3/4(k+2)} g(z_2 ) \jint \bra{\lambda _0} J^S (t)
\phi _{+}(z_2 )
\phi _{-}(z_1 ) \ket{\lambda _0} \\
& = & z^{3/4(k+2)} g(z_2 ) \jint \oint \frac{dw}{2\pi \sqrt{-1}}
\bra{\lambda _0} J^S (t)
[ \phi _{-}(z_2 ), J^- (w) ]_q
\phi _{-}(z_1 ) \ket{\lambda _0}. \nonumber
\end{eqnarray}
Thanks to the formulae of the OPE
given in Appendix A
we obtain
\begin{eqnarray}
f_1 (z_1 , z_2) & = & z^{3/4(k+2)} g(z_2 ) \jint \left\{
\oint_{q^{k+3}z_2 } \frac{dw}{2\pi \sqrt{-1}} \bra{\lambda _0} J^S (t)
\phi _{-}(z_2 ) J^- (w)
\phi _{-}(z_1 ) \ket{\lambda _0}
\right. \nonumber \\
& & \left. ~~~~~~~~-q
\oint_{q^{k+3}z_i , i=1,2} \frac{dw}{2\pi \sqrt{-1}}
\bra{\lambda _0} J^S (t) J^- (w)
\phi _{-}(z_2 )
\phi _{-}(z_1 ) \ket{\lambda _0}
\right\} \nonumber \\
& = & -q^{1+4/(k+2)} z^{3/4(k+2)} g(z_2 ) G(z_1 , z_2 )
\jint \varphi _1 (z_1 , z_2 , t),
\end{eqnarray}
where $G(z_1 , z_2 )$ comes from OPE of the $q$-vertex operators
\begin{eqnarray}
\phi _- (z_2 )\phi _- (z_1 ) & = &
G(z_1 , z_2 )
:\phi _- (z_2 )\phi _- (z_1 ):, \nonumber \\
G(z_1 , z_2 ) & = & (q^k z_2 )^{1/2(k+2)}
\prod_{m=1}^{\infty} \fra
{\inp{p^{m}z}{q^4 }\inp{p^{m}q^4 z}{q^4 }}
{\inp{p^{m}q^2 z}{q^4 }^2 }, \label{G}
\end{eqnarray}
while, the integrand of the Jackson integral is given as follows:
\begin{equation}
\varphi _1 (z_1 , z_2 , t)=qt^{-1-2/(k+2)}
\fra
{\inp{q p^2 z_1 /t}{p} \inp{qp^2 z_2 /t}{p}}
{\inp{q^{-1} p^2 z_1 /t}{p} \inp{q^{-1} pz_2 /t}{p}}.
\end{equation}

Let us check that
$f_1 (z_1, z_2 )$ depends upon $z=z_1 /z_2 $ only and
hence we may denote $f_1 (z_1 , z_2 )=f_1 (z)$.
Using the freedom of redefinition $t\mapsto z_2 t$
in the Jackson integral, we can rewrite
\begin{equation}
g(z_2 ) = -q^{-2-(k+8)/2(k+2)}z_{2}^{3/2(k+2)}
\left[ \jint t^{-1-2/(k+2)}
\fra{\inp{p^2 q/t}{p}}{\inp{pq^{-1}/t}{p}}\right]^{-1}\!\!\!\!\!\!\!,
 \label{g}
\end{equation}
\begin{equation}
\jint \varphi_1 (z_1 , z_2 , t) =
qz_{2}^{-2/(k+2)} \jint t^{-1-2/(k+2)}
\fra
{\inp{p^2 qz/t}{p} \inp{p^2 q/t}{p}}
{\inp{p^2 q^{-1} z/t}{p} \inp{pq^{-1}/t}{p}}.  \label{f1}
\end{equation}
Therefore we can regard $f_1 (z_1 , z_2 )$ as a function of $z$:
\begin{eqnarray}
f_1 (z) & = & z^{3/4(k+2)}\prod_{m=1}^{\infty} \fra
{\inp{p^{m}z}{q^4 }\inp{p^{m}q^4 z}{q^4 }}
{\inp{p^{m}q^2 z}{q^4 }^2 } \times \nonumber \\
&& \times \displaystyle\frac{\jint t^{-1-2/(k+2)}
\fra
{\inp{p^2 qz/t}{p} \inp{p^2 q/t}{p}}
{\inp{p^2 q^{-1}z/t}{p} \inp{pq^{-1}/t}{p}}}
{\jint t^{-1-2/(k+2)}
\fra{\inp{p^2 q/t}{p}}{\inp{pq^{-1}/t}{p}}}. \label{f}
\end{eqnarray}

Let us repeat the same argument with respect to $f_2 (z_1 , z_2)$.
\begin{eqnarray}
f_2 (z_1 , z_2)
& = & z^{3/4(k+2)} g(z_2 ) \jint \bra{\lambda _0} J^S (t)
\phi _- (z_2 ) \phi _+ (z_1 )
\ket{\lambda _0} \\
& = & z^{3/4(k+2)} g(z_2 ) \jint \oint \frac{dw}{2\pi \sqrt{-1}}
\bra{\lambda _0} J^S (t)
\phi _- (z_2 ) [ J^- (w), \phi _- (z_1 ) ]_q
\ket{\lambda _0}. \nonumber
\end{eqnarray}
Similarly we have
\begin{equation}
f_2 (z_1 , z_2)=
-q^{1+4/(k+2)} z^{3/4(k+2)} g(z_2 ) G(z_1 , z_2 )
\jint \varphi _2 (z_1 ,z_2 , t), \label{f2}
\end{equation}
where,
\begin{equation}
\varphi _2 (z_1 ,z_2 , t)=t^{-1-2/(k+2)}
\fra
{\inp{qp^2 z_1 /t}{p} \inp{qp z_2 /t}{p}}
{\inp{q^{-1}pz_1 /t}{p} \inp{q^{-1}pz_2 /t}{p}}. \label{p2}
\end{equation}
Thus we obtain
\begin{eqnarray}
f_2 (z) & = & q^{-1}z^{3/4(k+2)}\prod_{m=1}^{\infty} \fra
{\inp{p^{m}z}{q^4 }\inp{p^{m}q^4 z}{q^4 }}
{\inp{p^{m}q^2 z}{q^4 }^2 } \times \nonumber \\
&& \times \displaystyle\frac{\jint t^{-1-2/(k+2)}
\fra
{\inp{p^2 qz/t}{p} \inp{pq/t}{p}}
{\inp{pq^{-1}z/t}{p} \inp{pq^{-1}/t}{p}}}
{\jint t^{-1-2/(k+2)}
\fra{\inp{p^2 q/t}{p}}{\inp{pq^{-1}/t}{p}}}.
\end{eqnarray}
Note that $z^{-3/4(k+2)}f_{i}(z)$ is analytic around $z=0$ $(i=1,2)$.

\vskip 2mm
\noindent{\bf 5.2 $q$-KZ Equation} \quad
Now we show that
the $q$-difference system for $f_1 (z_1 , z_2), f_2 (z_1 , z_2)$
gives the $q$-KZ equation for two-point function.
Let us study the effect of $p$-shift
$z_1 \mapsto z_1 , z_2 \mapsto pz_2 $.
The change of $f_i (z_1 , z_2 )$
results from $z^{3/4(k+2)}$, $G(z_1 , z_2 )$
and $\phi _i (z_1 , z_2 , t)$. ($i=1, 2$.)
First $G(z_1 , z_2 )$ transforms as follows:
\begin{equation}
G(z_1 , pz_2 )=q^{1/2}\rho (pz) G(z_1 , z_2 ), \label{pG}
\end{equation}
where,
\begin{equation}
\rho (z)=q^{-1/2}
\fra{\inp{q^2 z}{q^4 }^2 }{\inp{qz}{q^4 }\inp{q^4 z}{q^4 }},
\end{equation}
is precisely the same factor appeared in
the image of the universal ${\cal R}$ matrix of
$\uqa$ \cite{FR}.
Next the contribution from the Jackson integral is given as follows:
(See Appendix C, as for detail.)
\begin{eqnarray}
\lefteqn{
\left(
\jint \varphi_{1}(z_{1},pz_{2},t),~~
\jint \varphi_{2}(z_{1},pz_{2},t) \right) } \nonumber \\
&& \qquad = \left(
\jint \varphi_{1}(z_{1},z_{2},t),~~
\jint \varphi_{2}(z_{1},z_{2},t) \right) \overline{R}(pz), \label{phi}
\end{eqnarray}
where,
\begin{equation}
\overline{R}(z)=\frac{1}{1-q^2 z}\left(
\begin{array}{cc}
1-z & q^{-1}-q \\
(q^{-3}-q^{-1})z & q^{-2}(1-z)
\end{array} \right), \label{Rcheck}
\end{equation}
is just the zero-weight part of the $R$ matrix of the six vertex model
up to a similarity transformation.

By combining eqs. (\ref{pG}), (\ref{phi}), (\ref{Rcheck}),
and the factor from $z^{3/4(k+2)}$ we obtain
\begin{equation}
\left( \begin{array}{l}
        f_1 (pz) \\
        f_2 (pz)
\end{array} \right)
=  \fra{\rho (pz)}{1-pq^2 z}
\left( \begin{array}{cc}
        q^2 (1-pz) & pq^{-1}(1-q^{2})z \\
        q(1-q^{2}) & (1-zp)
\end{array} \right)
\left( \begin{array}{l}
        f_1 (z) \\
        f_2 (z)
\end{array} \right).
\end{equation}
It coincides with the $q$-KZ equation \cite{FR}
for the two-point function.

This recursion formula implies
$$
qf_1 (pz)+f_2 (pz)
=\rho (pz)(qf_1 (z)+f_2 (z)). \label{qf}
$$
Compare the coefficients of $z^{3/4(k+2)}$
of both sides of (\ref{qf}), then we obtain
$$
qf_1 (z)+f_2 (z)=0.
$$
Therefore we have $q$-difference equation of the first order
\begin{eqnarray}
\fra{f_1 (pz)}{f_1 (z)} & = & q^{3/2}
\fra{\inp{pq^{-2}z}{q^4 }\inp{pq^6 z}{q^4 }}
    {\inp{pqz}{q^4 }\inp{pq^4 z}{q^4 }}, \\
f_2 (z) & = & -qf_1 (z).
\end{eqnarray}
In particular if we put $k=1$,
by solving the above $q$-difference equation we have
\begin{equation}
\Psi (z_1 , z_2 )=z^{1/4}\fra{\inp{q^6 z}{q^4 }}{\inp{q^4 z}{q^4 }}
(v_{+}\otimes v_{-}-qv_{-}\otimes v_{+}), \label{2pt}
\end{equation}
which reproduces the known results\footnote
{We use the opposite ordering of two $V$s to that of ref. \cite{DFJMN}.}.

\section{Conclusion}

In this paper we discuss a bosonization of $q$-vertex operator on the
basis of the Fock representation of $\uqa$. We propose an integral
formula for $N$-point functions of the $q$-vertex operators with the
help of the screening charges.  Matsuo \cite{Ma2} and Reshetikhin
\cite{Re} have obtained integral formulae from the viewpoint of the
$q$-KZ equation.  The relations among these three integral formulae
should be clarified.

After performing all the residue calculus of two-point function, we have
Jackson integral of Jordan-Pochhammer type \cite{Mi} \cite{Ma1}. It is
intriguing that the scalar factor which arises in the image of the
universal $\cal R$ matrix naturally appears in the OPE of elementary
vertex operators.

We would like to check all the intertwining properties of the elementary
$q$-vertex operators for general case.  The analogy with CFT is quite
remarkable; we can deform $\slth$ currents, screening current, and
vertex operators \`{a} la Tsuchiya-Kanie \cite{TK}. However, we have no
counterpart of Virasoro algebra, and the meaning of the spectral
parameters of the $q$-vertex operators is not yet obvious.

Recently Matsuo \cite{Ma3} constructed another bosonization of $\uqap$.
It is interesting to investigate the connection between his bosonization
and ours.
After completing this work we received a preprint by
Abada et al. \cite{ABG}.

\section*{Acknowledgement}
The authors thank T. Eguchi, M. Jimbo, A. Matsuo, T. Miwa and K.
Sugiyama for helpful discussions and useful comments.  They also thank
H. Awata, T. Inami, T. Nakatsu, A. Nakayashiki, T. Sano and Y. Yamada
for encouragement and their interest in this work.  J.S. is grateful to
kind hospitality at YITP and RIMS.

This work was supported in part by the Grant-in-Aid for Scientific Research
from the Ministry of Education, Science and Culture
 (No. 04245206 and No. 04-2297).

\appendix

\section{Operator Product Expansion Formulae}

In this appendix we list the operator product expansion formulae among
the $\uqa$-current $J^{-}(z)$, the screening current $J^{S}(z)$ and the
elementary $q$-vertex operator $\phi^{(l)}_{l}(z)$.

We split $J^{-}(z)$ into two parts
\begin{eqnarray*}
   J^{-}(z) &\equiv& \fra{1}{(q-q^{-1})z}
       \left\{ \jup (z) - \jdn (z) \right\}, \\
   \jup (z) &=& : \exp \left\{
      \bosa{k+2}{q^{k}z}{-\fra{k+2}{2}}
      - \bosa{k+2}{q^{-2}z}{\fra{k+2}{2}} \right. \\
            & & \quad \qquad \left.
      + \bosb{2}{z}{-1}
      + \bosc{2}{q^{-1}z}{0}
       \right\} : ,
\\
   \jdn (z) &=& : \exp \left\{
      \bosa{k+2}{q^{-k-4}z}{-\fra{k+2}{2}}
      - \bosa{k+2}{q^{-2}z}{\fra{k+2}{2}} \right. \\
            & & \quad \qquad \left.
      + \bosb{2}{q^{- 2 k -4}z}{-1}
      + \bosc{2}{q^{-2k-3}z}{0}
       \right\} : .
\end{eqnarray*}

Similarly, we put
\begin{eqnarray*}
   J^{S}(z) &\equiv&
   {-} \fra{1}{(q-q^{-1})z} \left\{ \jsup (z) - \jsdn (z) \right\},
\\
   \jsup (z) &=& : \exp \left\{
      - \bosa{k+2}{q^{-2}z}{-\fra{k+2}{2}}
      - \bosb{2}{q^{-k-2}z}{-1}
      - \bosc{2}{q^{-k-1}z}{0}
       \right\} : ,
\\
   \jsdn (z) &=& : \exp \left\{
      - \bosa{k+2}{q^{-2}z}{-\fra{k+2}{2}}
      - \bosb{2}{q^{-k-2}z}{-1}
      - \bosc{2}{q^{-k-3}z}{0}
       \right\} : .
\end{eqnarray*}


\begin{equation}
   \begin{array}{rcll}
      \jup(z) \phi^{(l)}_{l}(w) &=&
	 \fra{q^{l} z - q^{k+2} w}{z-q^{l+k+2} w}
	 : \jup(z) \phi^{(l)}_{l}(w) :
	 & \quad |z|>q^{k+2-l}|w|
      \hh \\
      \jdn(z) \phi^{(l)}_{l}(w) &=&
	 q^{-l}
	 : \jdn(z) \phi^{(l)}_{l}(w) :
	 & \quad
      \hh \\
      \phi^{(l)}_{l}(w) \jup(z) &=&
	 : \phi^{(l)}_{l}(w) \jup(z) :
	 & \quad
      \hh \\
      \phi^{(l)}_{l}(w) \jdn(z) &=&
	 \fra{w - q^{-l-k-2} z}{w - q^{l-k-2} z}
	 : \phi^{(l)}_{l}(w) \jdn(z) :
	 & \quad |w|>q^{-l-k-2}|z|
   \end{array}
\end{equation}


\begin{equation}
   \begin{array}{rcll}
      \jsup(w) \jup(z) &=&
	 q^{-1}
	 : \jup(z) \jsup(w) :
	 & \quad
      \hh \\
      \jsup(w) \jdn(z) &=&
	 \fra{q^{-1} w - q^{-k-1} z}{w - q^{-k-2} z }
	 : \jdn(z) \jsup(w) :
	 & \quad |w| > q^{-k-2} |z|
      \hh \\
      \jsdn(w) \jup(z) &=&
	 \fra{q w - q^{k+1} z}{w-q^{k+2}z}
	 : \jup(z) \jsdn(w) :
	 & \quad |w| > q^{k} |z|
      \hh \\
      \jsdn(w) \jdn(z) &=&
	 q
	 : \jdn(z) \jsdn(w) :
	 & \quad
      \hh
   \end{array}
\end{equation}
\begin{equation}
   \begin{array}{rcll}
      \jup(z) \jsup(w) &=&
	 q^{-1}
	 : \jup(z) \jsup(w) :
	 & \quad
      \hh \\
      \jup(z) \jsdn(w) &=&
	 \fra{q^{-1} z - q^{-k-1} w}{z - q^{-k-2} w}
	 : \jup(z) \jsdn(w) :
	 & \quad |z| > q^{-k-2} |w|
      \hh \\
      \jdn(z) \jsup(w)  &=&
	 \fra{q z - q^{k+1} w}{ z - q^{k+2}w}
	 : \jdn(z) \jsup(w) :
	 & \quad |z| > q^{k} |w|
      \hh \\
      \jdn(z) \jsdn(w) &=&
	 q
	 : \jdn(z) \jsdn(w) :
	 & \quad
   \end{array}
\end{equation}


\begin{equation}
   \begin{array}{rcll}
      \jsup(z) \phi^{(l)}_{l}(w) &=&
	 \fra{\inp{q^{l} p w/z}{p}}{\inp{q^{-l} p w/z}{p}}
	 ( q^{-2} z )^{-l/2(k+2)}
	 : \jsup(z) \phi^{(l)}_{l}(w) :
	 & \quad |z| > q^{-l} p |w|
      \hh \\
      \jsdn(z) \phi^{(l)}_{l}(w) &=&
	 \fra{\inp{q^{l} p w/z}{p}}{\inp{q^{-l} p w/z}{p}}
	 ( q^{-2} z )^{-l/2(k+2)}
	 : \jsdn(z) \phi^{(l)}_{l}(w) :
	 & \quad |z| > q^{-l} p |w|
      \hh \\
      \phi^{(l)}_{l}(w) \jsup(z) &=&
	 \fra{\inp{q^{l} w/z}{p}}{\inp{q^{-l} w/z}{p}}
	 ( q^{k} w )^{-l/2(k+2)}
	 : \jsup(z) \phi^{(l)}_{l}(w) :
	 & \quad |w| > q^{-l} |z|
      \hh \\
      \phi^{(l)}_{l}(w) \jsdn(z) &=&
	 \fra{\inp{q^{l} w/z}{p}}{\inp{q^{-l} w/z}{p}}
	 ( q^{k} w )^{-l/2(k+2)}
	 : \jsdn(z) \phi^{(l)}_{l}(w) :
	 & \quad |w| > q^{-l} |z|
   \end{array}
\end{equation}


\begin{equation}
  \phi^{(l)}_{l}(z)  \phi^{(l)}_{l}(w) =
  (q^{k}z)^{l^{2}/2(k+2)}
  \prod_{m=1}^{\infty} \fra{\inp{p^{m}q^{2(1-l)}w/z}{q^{4}}
    \inp{p^{m}q^{2(1+l)}w/z}{q^{4}}}
   {\inp{p^{m}q^{2}w/z}{q^{4}}^{2}}
  : \phi^{(l)}_{l}(z)  \phi^{(l)}_{l}(w) :
\end{equation}

\section{Normalization of $q$-Vertex Operators}

Here we consider the normalization of the following VOs
\begin{equation}
\begin{array}{lllll}
\tilde{\Phi}^{\lambda_1\;V^{(1)}}_{\lambda_0}(z) & : &
V(\lambda_0) & \rightarrow & V(\lambda_1) \otimes V^{(1)}_{z} , \hh \\
\tilde{\Phi}^{\lambda_0\;V^{(1)}}_{\lambda_1}(z) & : &
V(\lambda_1) & \rightarrow & V(\lambda_0) \otimes V^{(1)}_{z} . \hh
\end{array}
\end{equation}
These VOs have the following leading terms
\begin{equation}
\begin{array}{l}
\tilde{\Phi}^{\lambda_1\;V^{(1)}}_{\lambda_0}(z) \ket{\lambda_{0}} =
\ket{\lambda_{1}} \otimes v_{-} + \cdots , \hh \\
\tilde{\Phi}^{\lambda_0\;V^{(1)}}_{\lambda_1}(z) \ket{\lambda_{1}} =
\ket{\lambda_{0}} \otimes v_{+} + \cdots . \hh
\end{array}
\end{equation}

These VOs are bosonized as
\begin{equation}
\begin{array}{rl}
\tilde{\Phi}^{\lambda_1\;V^{(1)}}_{\lambda_0}(z) & =
g^{\lambda_1\;V^{(1)}}_{\lambda_0}(z) \left[
\phi_{+}(z) \otimes v_{+} + \phi_{-}(z) \otimes v_{-} \right], \hh \\
\tilde{\Phi}^{\lambda_0\;V^{(1)}}_{\lambda_1}(z) & =
g^{\lambda_0\;V^{(1)}}_{\lambda_1}(z) \left[
\jint J^{S}(t) \; \phi_{+}(z) \otimes v_{+} +
\jint J^{S}(t) \; \phi_{-}(z) \otimes v_{-} \right] . \hh
\end{array}
\end{equation}

We can get these normalization functions
$g^{\lambda_1\;V^{(1)}}_{\lambda_0}(z), g^{\lambda_0\;V^{(1)}}_{\lambda_1}(z)$
by calculating the leading term explicitly.

First we have
\begin{equation}
\begin{array}{rl}
  & \phi_{-}(z) \ket{0,0} \hh \\
= & :\exp \left\{ \bosal{1}{2}{k+2}{q^{k}z}{\fra{k+2}{2}} \right\}:
    \ket{0,0} \hh \\
= & \ket{1/2,0} + \cdots , \hh
\end{array}
\end{equation}
then we get
\begin{equation}
g^{\lambda_1\;V^{(1)}}_{\lambda_0}(z)=1.
\end{equation}

Next we can see
\begin{equation}
\begin{array}{rl}
  & \jint J^{S}(t) \; \phi_{+}(z) \ket{1/2,0} \hh \\
= & \jint J^{S}(t) \int \fra{dw}{2 \pi \sqrt{-1}}
    \left[ \phi_{-}(z) , J^{-}(w) \right]_{q}
    \exp \left\{ \fra{Q_{a}}{2(k+2)} \right\} \ket{0,0} \hh \\
= & -q^{2+(k+8)/2(k+2)}z^{1/2(k+2)}
    \jint t^{-1-2/(k+2)}\fra{\inp{qp^{2}z/t}{p}}{\inp{q^{-1}pz/t}{p}}
    \ket{0,0} + \cdots , \hh
\end{array}
\end{equation}
likewise. Then in this case
\begin{equation}
   g^{\lambda_0\;V^{(1)}}_{\lambda_1}(z)
=  -q^{-2-(k+8)/2(k+2)}z^{-1/2(k+2)}
\left[
  \jint t^{-1-2/(k+2)}\fra{\inp{qp^{2}z/t}{p}}{\inp{q^{-1}pz/t}{p}}
\right]^{-1}
\end{equation}
holds.

\section{Difference Properties of Jackson Integrals}

In this appendix, we calculate the difference system satisfied by
the following functions:
\begin{equation}
\jint{\varphi_{1}(z_{1},z_{2},t)}, \qquad \jint{\varphi_{2}(z_{1},z_{2},t)},
\end{equation}
where $\varphi_{1}(z_{1},z_{2},t)$,
and $\varphi_{2}(z_{1},z_{2},t)$ are given as
\begin{equation}
\begin{array}{rl}
\varphi_{1}(z_{1},z_{2},t) =&
q\fra{\inp{qp^{2}z_{1}/t}{p} \inp{qp^{2}z_{2}/t}{p}}
     {\inp{q^{-1}p^{1}z_{1}/t}{p} \inp{q^{-1}p^{2}z_{2}/t}{p}}
t^{-1-2/(k+2)} , \hh \\
\varphi_{2}(z_{1},z_{2},t) =&
 \fra{\inp{qpz_{1}/t}{p} \inp{qp^{2}z_{2}/t}{p}}
     {\inp{q^{-1}pz_{1}/t}{p} \inp{q^{-1}pz_{2}/t}{p}}
t^{-1-2/(k+2)} . \hh
\end{array}
\end{equation}
We note that these functions are the Jackson integrals of
Jordan-Pochhammer type. For the general theory of the difference system
for the Jackson integrals of Jordan-Pochhammer type, we refer the reader
to \cite{Mi}\cite{Ma1}.

To find the difference equation, we use the following identity:
\begin{equation}
\jint \varphi_{i}(z_{1},z_{2},t) =
\jint p \varphi_{i}(z_{1},z_{2},pt) .
\end{equation}
Since we have
\begin{equation}
\begin{array}{rl}
p \varphi_{1}(z_{1},pz_{2},pt) =&
p \varphi_{1}(z_{1}, z_{2},pt) \fra{1-pz}{1-pq^{2}z} +
  \varphi_{2}(z_{1}, z_{2}, t) \fra{(1-q^{2})pq^{-3}z}{1-pq^{2}z} , \hh \\
p \varphi_{2}(z_{1},pz_{2},pt) =&
p \varphi_{1}(z_{1}, z_{2},pt) \fra{(1-q^{2})q^{-1}}{1-pq^{2}z} +
  \varphi_{2}(z_{1}, z_{2}, t) \fra{(1-zp)q^{-2}}{1-pq^{2}z} , \hh
\end{array}
\end{equation}
we get the following difference equation:
\begin{eqnarray}
\lefteqn{
\left( \begin{array}{cc}
        \jint \varphi_{1}(z_{1},pz_{2},t), &
        \jint \varphi_{2}(z_{1},pz_{2},t)
\end{array} \right)
 } \nonumber \\
&& =
\left( \begin{array}{cc}
        \jint \varphi_{1}(z_{1},z_{2},t),  &
        \jint \varphi_{2}(z_{1},z_{2},t)
\end{array} \right)
\left( \begin{array}{cc}
  \fra{1-pz}{1-pq^{2}z}            & \fra{(1-q^{2})q^{-1}}{1-pq^{2}z} \hh \\
  \fra{(1-q^{2})pq^{-3}z}{1-pq^{2}z} & \fra{(1-zp)q^{-2}}{1-pq^{2}z}
\end{array} \right). \nonumber \\
\end{eqnarray}

\end{document}